\documentclass[12pt]{article}
\usepackage{amsmath}
\usepackage{amssymb}
\usepackage{citesort}
\usepackage{graphicx}
\usepackage[nomarkers]{endfloat}
\usepackage{pstricks}
\numberwithin{equation}{section}
\begin{document}
\setcounter{page}{0}
\thispagestyle{empty}
\begin{flushright}
\end{flushright}
\vspace*{2.5cm}
\begin{center}
{\large\bf Charged Domain Walls}
\end{center}
\vspace*{2cm}
\renewcommand{\thefootnote}{\fnsymbol{footnote}}
\begin{center}
L. Campanelli$^{1,2,}$\protect\footnote{Electronic address: {\tt
Leonardo.Campanelli@ba.infn.it}},
P. Cea$^{1,2,}$\protect\footnote{Electronic address: {\tt
Cea@ba.infn.it}},
G.~L. Fogli$^{1,2,}$\protect\footnote{Electronic address: {\tt
Fogli@ba.infn.it}} and
L. Tedesco$^{1,2,}$\protect\footnote{Electronic address: {\tt
Luigi.Tedesco@ba.infn.it}} \\[0.5cm] $^1${\em Dipartimento di Fisica,
Universit\`a di Bari, I-70126 Bari, Italy}\\[0.3cm] $^2${\em INFN
- Sezione di Bari, I-70126 Bari, Italy}
\end{center}
\vspace*{0.5cm}
\begin{center}
{July, 2003}
\end{center}
\vspace*{1.0cm}
%
%
%
%
%
%
\renewcommand{\abstractname}{\normalsize Abstract}
\begin{abstract}
In this paper we investigate Charged Domain Walls (CDW's),
topological defects that acquire surface charge density $Q$
induced by fermion states localized on the walls. The presence of
an electric and magnetic field on the walls is also discussed. We
find a relation in which the value of the surface charge density
$Q$ is connected with the existence of such topological defects.
\end{abstract}
%
%
%
%
\vspace*{0.5cm}
\begin{flushleft}
\end{flushleft}
\vfill
\newpage
%
%
%
%
%
%
\renewcommand{\thesection}{\normalsize{\arabic{section}.}}
\section{\normalsize{Introduction}}
\renewcommand{\thesection}{\arabic{section}}
In the fundamental theories of elementary particles, the
spontaneous symmetry breaking by the Higgs mechanism play a
central role. Such theories in general have  a degenerate vacuum
manifold with a non-trivial topology. In the recent years a lot of
investigations on topological defects have been done, with
applications on different fields, as for instance particle physics
or cosmology \cite{VILENKIN}. In this paper we study topological
defects, called domain walls, that are inherent to field theories
with spontaneously broken discrete symmetries. The scalar field
has isolated minima and the walls are surfaces interpolating
between minima of the scalar potential with different vacuum
expectation values of the scalar field. This phenomenon,
occurrence of the domain walls, is quite common in solid state
physics. In cosmology, such structures can form by the Kibble
mechanism \cite{KIBBLE}, and therefore must be taken into account
in cosmological considerations.
\\
The idea of this work is connected with the so called Q-balls.
These non topological solitons can be created during  a phase
transition \cite{GELMINI}. In the mid-70's new topological
solitons were considered in the context of systems with two scalar
fields \cite{LEE}. In the mid-80's they reappeared Coleman's works
\cite{COLEMAN}, and they were given the name of Q-balls. In the
mid-90's, the Q-balls were connected with models of supersymmetry
breaking, as in Ref.~\cite{{KUSENKO1},{KUSENKO2}}, in which the
Q-balls are condensates of squark or slepton particles. These
condensates are connected in baryogenesis by the so-called
Affleck-Dine mechanism \cite{A-D}. Thus, the Q-balls might be
important because they might contribute to the dark matter in the
Universe \cite{{KUSENKO2},{KUSENKO3}}. As a consequence, the
people began to consider Q-balls and Q-strings (extended objects
of the Q-balls type \cite{{BATTYE},{AXENIDES}}). The study of
Q-walls and Q-strings is also connected with their lifetime
\cite{MCKENZIE}, that is arbitrarily large. This is a very
important point because these configurations might be of relevance
in the formation of structures in the early Universe. Moreover, it
has been considered domain walls carrying a U(1) charge. These
system are composed by two interacting scalar fields, the Higgs
real field and a complex scalar field \cite{{LENSKY,FRIEDBERG}}.
\\
It is important to consider two aspects in this analysis. The
first is the interaction between particles and domain walls. This
study start from the Voloshin's paper \cite{VOLOSHIN} and
continues with the work of Ref.~\cite{FARRAR}. A recent analysis
by the authors of Ref.~\cite{CEA,CEA2,AYALA}, considers the
scattering of fermions off domain walls at the electroweak phase
transition in presence of a primordial magnetic field. The second
paper deals with the trapping of the massless fermions in these
defects. In fact it is well known that Dirac fermions with Yukawa
coupling with scalar field develop zero mode solution near a
domain wall \cite{{JACKIW},{SEMENOF}}. These zero mode behave like
massless fermions in the two dimensional space of the wall.
Recently we have analyzed the zero energy solutions localized on
the wall in presence of a magnetic field on the wall. The
localized states are a peculiar characteristic of the domain wall
and they play an important role in the dynamics of the walls. The
localizing of the fermions in the core of the defect is important
for building our analysis in this paper in which we investigate
charged domain walls (CDW), in which the charge is not due to the
charged scalar fields, as in the Q-balls theory, but to the
interaction with the massless fermions. In particular our analysis
connects the charge $Q$ with some parameters of the theory in
order to have some constraint about the existence of such CDW's.
The physical motivation to investigate CDW's was provided mainly
through the works by Lopez \cite{LOPEZ} and Gr\o n \cite{GRON} and
the study of repulsive gravitational fields \cite{AMUNDSEN}.
\\
The organization of this paper is as follows. In the next section
we study the equation of motion of lagrangian density of a
$\lambda \phi^4$ theory with the  scalar field coupled by Yukawa
coupling to massless fermions. In particular we find the wave
function of localized zero mode solution. In Sec.~III we also
introduce the electromagnetic field and we find the general
expression for the wave function interacting with domain wall.
Moreover we analyze the magnetic and electric field near the wall.
In Sec.~IV we study the system formed by the domain wall and a
fermionic plane wave, in order to compare the energy density of
this system with the system in Sec.~III. In Sec.~V we conclude
with some cosmological implications of these charged domain walls.
%
%
%
%
%
%
%
\renewcommand{\thesection}{\normalsize{\arabic{section}.}}
\section{\normalsize{Fermionic zero-modes localized on the wall}}
\renewcommand{\thesection}{\arabic{section}}
In this Section we  consider a single real self-interacting scalar
field, coupled with a massless Dirac fermion trough  Yukawa
coupling. The Lagrangian density of the system is
\begin{equation}
{\cal {L}} = \frac{1}{2} \partial_{\mu} \phi \, \partial^{\mu}
             \phi - V(\phi) + \bar{\psi} \,
             (i \, / \!\!\!\partial - g_Y \phi) \, \psi, \;\;\;\;
             V(\phi) = \frac{\lambda}{4} (\phi^2 - \eta^2)^2,
\end{equation}
where the potential $V(\phi)$ has degenerate minima at $\phi=\pm
\eta$. In this paper we restrict ourselves to the
$(1+1)$-dimensional case, and we suppose that
\begin{equation}
\phi = \phi(x),
\end{equation}
\begin{equation}
\psi = \psi(x,t) = \xi(x) \, e^{-i \omega t},
\end{equation}
where $\phi$ and $\xi$ are real functions of $x$. We will use the
following representation of the Dirac matrices,
\begin{equation}
\gamma^0 =
          \begin{pmatrix} \sigma^3 & 0 \\ 0 & -\sigma^3 \end{pmatrix}\!\!, \;\;
\gamma^1 =
          \begin{pmatrix} i\sigma^1 & 0 \\ 0 &
          - i\sigma^1 \end{pmatrix}\!\!, \;\;
\gamma^2 =
          \begin{pmatrix} i\sigma^2 & 0 \\ 0 &
          - i\sigma^2 \end{pmatrix}\!\!, \;\;
\gamma^3 =
\begin{pmatrix}0 & i \\ i & 0\end{pmatrix} \!\! ,
\end{equation}
where $\sigma^i$, $i=1,2,3$, are the Pauli matrices. The
Lagrangian (2.1) with Eqs.~(2.2)-(2.4) implies the equations of
motion
\begin{equation}
\phi'' + \lambda \, (\phi^2 - \eta^2) \, \phi =
                                              g_Y \, {\bar{\xi}} \xi,
\end{equation}
\begin{equation}
i \gamma^1 \xi' + (\omega \gamma^0 - g_Y \phi) \, \xi = 0.
\end{equation}
(Here, and throughout, a prime will denote differentiation with
respect to $x$.) In this paper we make the following ansatz for
the spinor $\xi$,
\begin{equation}
\xi = \frac{1}{\sqrt{2}} \left(
\begin{array}{c}
u
\\
u
\\
v
\\
v
\end{array}
\right) \!\! ,
\end{equation}
where $u$ and $v$ are real scalar functions of $x$. Because ${\bar
{\xi}} \xi=0$, Eq.~(2.5) take the form
\begin{equation}
\phi'' - \lambda \, (\phi^2 - \eta^2) \, \phi = 0,
\end{equation}
that is the equation of motion for $\lambda \, \phi^4$ theory. It
is simple to see that Eq.~(2.8) admits the solution describing the
transition between two adjacent regions with different values of
$\phi$, that is $\phi=+\eta$ and $\phi=-\eta$. The profile takes
the form
\begin{equation}
\phi(x) = \eta \, \tanh (x/\Delta),
\end{equation}
where $\Delta= \sqrt{2/\lambda}\,\eta^{-1 }$ is the thickness of
the wall, and gives origin to the so called ``domain wall" which
is thought to be formed in a continuous phase transition by the
Kibble mechanism \cite{KIBBLE}; the domain wall is
the interpolating region of rapid change of the scalar field. \\
With the ansatz (2.7), Eq.~(2.6) splits in two equations
\begin{equation}
u' + g_Y \, \phi \, u = 0,
\end{equation}
\begin{equation}
v' - g_Y \, \phi \, v = 0,
\end{equation}
with $\omega=0$. The physical solution of Eq.~(2.11) is $v=0$,
since the solution $v \neq 0$ is not localized near the wall;
moreover Eq.~(2.10) gives us the following solution:
\begin{equation}
u(x) = N \, {\left[\mbox{sech}(x / \Delta) \right]}^{m_{0}\Delta},
\end{equation}
where $N$ is a normalization constant, and $m_{0}=g_Y \eta$ is the
fermion mass in the broken phase. Finally, the expression for
$\psi$ is
\begin{equation}
\psi(x)= \frac{N}{\sqrt{2}} \,
         {\left[ \mbox{sech} (x / \Delta)
         \right]}^{m_{0}\Delta}
                  \left(
                        \begin{array}{c}
                            1 \\ 1 \\ 0 \\ 0
                        \end{array}
                  \right) \!\! .
\end{equation}
The normalization constant $N$ is bound to the surface charge
density $Q$ by
\begin{equation}
Q = \int_{- \infty}^{+\infty} \! dx \, j^0 =
    \int^{+\infty}_{-\infty} \! dx \,|\psi|^2,
\end{equation}
and take the following value,
\begin{equation}
N^2 = \frac{Q}{\Delta \, B(m_{0}\Delta,1/2)} \, ,
\end{equation}
where $B(x,y)$ is the Bernoulli Beta Function.
%
%
%
%
%
%
\renewcommand{\thesection}{\normalsize{\arabic{section}.}}
\section{\normalsize{Topological charged domain walls}}
\renewcommand{\thesection}{\arabic{section}}
In this Section we perform a more complete analysis by considering
massless Dirac fermion coupled to a real scalar field trough the
Yukawa coupling in presence of the electromagnetic field generated
by fermions. Explicitly the dynamics is determined by the
Lagrangian density
\begin{equation}
{\cal {L}} = \frac{1}{2} \partial_{\mu} \phi \,
             \partial^{\mu} \phi - \frac{\lambda}{4}
             (\phi^2 - \eta^2)^2 + \bar{\psi} \,
             (i \, / \!\!\!\partial - g_Y \phi - e \,
             \slash\!\!\!\!\!\:\!A) \, \psi -
             \frac{1}{4} F_{\mu \nu} F^{\mu \nu}.
\end{equation}
Let us suppose that
\begin{equation}
\phi = \phi(x),
\end{equation}
\begin{equation}
\psi = \psi(x,t) = \xi(x) \, e^{-i \omega t},
\end{equation}
\begin{equation}
A^{\mu} = A^{\mu}(x),
\end{equation}
where $\phi$, $\xi$ and $A^{\mu}$ are real functions of $x$. We
use the Lorentz gauge for the electromagnetic field,
$\partial_{\mu} \, A^{\mu} = 0$. The Lagrangian density (3.1) with
Eqs.~(3.2)-(3.4) implies the equations of motion
\begin{equation}
\phi'' - \lambda \, (\phi^2 - \eta^2) \, \phi =
                        g_Y \, {\bar{\xi}} \xi,
\end{equation}
\begin{equation}
i \gamma^1 \xi' + (\omega \gamma^0 - g_Y \phi - e \,
                  \slash\!\!\!\!\!\:\!A) \, \xi = 0,
\end{equation}
\begin{equation}
(A^{\mu})'' = - e \, {\bar{\xi}} \gamma^{\mu} \xi.
\end{equation}
With the ansatz (2.7), Eqs.~(3.5)-(3.7) become
\begin{equation}
\phi'' + \lambda \, (\phi^2 - \eta^2) \, \phi = 0,
\end{equation}
\begin{equation}
u' + g_Y \, \phi \, u = 0,
\end{equation}
\begin{equation}
v' - g_Y \, \phi \, v = 0,
\end{equation}
\begin{equation}
(A^0)'' = (A^2)'' = - 2 \, e \, (u^2 + v^2),
\end{equation}
\begin{equation}
(A^1)'' = (A^3)'' = 0,
\end{equation}
with $eA^0 = eA^2 + \omega$. Without lost of generality we can put
$A^1=A^3=0$. The equation (3.8) has been already analyzed in
previous Section. The solutions of Eqs.~(3.9) and (3.10) are given
by Eq.~(2.12) and $v=0$ as observed in previous Section. Therefore
the expression for $\psi$ is
\begin{equation}
\psi(x,t) = \frac{N}{\sqrt{2}} \;
          {\left[ \mbox{sech} (x / \Delta)
          \right]}^{m_{0}\Delta}
                     \left(
                           \begin{array}{c}
                                 1 \\ 1 \\ 0 \\ 0
                           \end{array}
                     \right) e^{-i \omega t},
\end{equation}
where $N$ is given by Eq.~(2.15). Taking into account Eq.~(3.11)
we obtain the electric and magnetic field $\textbf{E} =
(E_x,0,0)$, $\textbf{B} = (0,0,B_z)$, where
\begin{equation}
E_x = -(A^0)' = 2e \int_0^x \!\! dx' \,[u^2(x') + v^2(x')] =
                         e \int_0^x \!\! d x' |\psi(x',t)|^2,
\end{equation}
and $B_z = -E_x$. The electric field is perpendicular to the wall,
while the magnetic field is parallel to the wall. They grow with
$x$ where the domain wall is at $x=0$. The maximum value of the
electric field is obtained when $x \rightarrow + \infty$,
therefore
\begin{equation}
E^{(max)}_x = e \int^{+ \infty}_0 \!\! dx' \, |\psi(x',t)|^2 =
              \frac{e\, Q}{2}.
\end{equation}
In the same way $B^{(max)}_z = e\, Q/2$. It is important to remark
that this behavior is strictly connected to the non realistic case
of infinite domain wall. Indeed, if we consider a disk of radius
$R$ in which a charge density $eQ$ is stored, the value of the
electric field is
$E_x =(eQ/2)(1-|x|/\sqrt{R^2+x^2})$,
where $x$ is the distance from the disk: the electric field
reduces when we depart from the disk. Therefore for finite domain
walls, we expect that the electric and magnetic fields are null in
the core of the defect, they reach the maximum value on the edge
of the wall and then they decrease when we depart from the wall.
\\
Let us consider a charged domain wall, that is a wall with fermion
states localized on it. The surface energy density of such
configuration is
\begin{equation}
\sigma_{T} = \sigma_w + \sigma_{\psi} + \sigma_{E} + \sigma_{B},
\end{equation}
where the density $\sigma_w$ of the kink (2.9) is
\begin{equation}
\sigma_w = \frac{8}{3 \lambda \Delta^3},
\end{equation}
the energy density referred to $\psi$ is
\begin{equation}
\sigma_{\psi} = \omega \, Q,
\end{equation}
while the energy densities associated to magnetic and electric
fields are
\begin{equation}
\sigma_B = \sigma_E = \int^{\infty}_{-\infty} \!\! dx \,
                      \frac{E^2_x}{2}.
\end{equation}
In analogy with the case of a charged disk of finite radius, we
can suppose that
\begin{equation}
\int^{+\infty}_{-\infty} \!\! dx \, E^2_x \;
 \sim \, E_x^{(max)} \, l = e^2 \, Q^2 \, l,
\end{equation}
with $l$  is the linear dimension of the wall.
%
%
%
%
%
%
\renewcommand{\thesection}{\normalsize{\arabic{section}.}}
\section{\normalsize{Non topological charged domain walls}}
\renewcommand{\thesection}{\arabic{section}}
In this Section we will consider the following physical system: a
domain wall and a fermionic  plane wave  with his electromagnetic
field in order to compare the energy density of such configuration
with the physical case of the previous Section. Let us consider
the equations of motion (3.5)-(3.7) in the case in which
\begin{equation}
\phi = \phi_{vac} + \delta \phi,
\end{equation}
\begin{equation}
\psi = \psi_{vac} + \delta \psi,
\end{equation}
\begin{equation}
A^{\mu} = A^{\mu}_{vac} + \delta A^{\mu},
\end{equation}
where the vacuum states of the system are
\begin{equation}
\phi_{vac} = \pm \eta, \;\;\; \psi_{vac}=0, \;\;\;
             A^{\mu}_{vac}=0.
\end{equation}
In other words we write the field in terms of the fluctuations
over the vacuum background. Under these assumptions, to the first
order, the equations of motion in the linear approximation are
\begin{equation}
\partial_{\mu} \partial^{\mu} \delta \phi +
         2 \lambda \, \eta^2 \delta \phi=0,
\end{equation}
\begin{equation}
(i \gamma^{\mu} \partial_{\mu} - g_Y \phi_{vac}) \, \delta \psi=0,
\end{equation}
\begin{equation}
\partial_{\mu} \partial^{\mu} \delta A^{\nu}=
e \, \delta \bar{\psi} \, \gamma^{\nu} \delta \psi .
\end{equation}
Remembering that we have respectively $\phi_{vac}=+\eta$ for $x>0$
and $\phi_{vac}=-\eta$ for $x<0$, at fixed charge of
configuration, the solution of Eqs.~(4.5)-(4.7), that minimized
the deviation of energy from its vacuum value, is
\begin{equation}
\delta \phi = 0,
\end{equation}
\begin{equation}
\delta \psi_- = \frac{C}{\sqrt{2}}
                   \left(
                         \begin{array}{c}
                            1 \\ 0 \\ 0 \\ 1
                         \end{array}
                   \right)
                e^{i \, m_0 t}  \;\;\;\; \mbox{if $x<0$},
\end{equation}
\begin{equation}
\delta \psi_+ = \frac{C}{\sqrt{2}}
                   \left(
                         \begin{array}{c}
                            0 \\ 1 \\ 1 \\ 0
                         \end{array}
                   \right)
                e^{i \, m_0 t} \;\;\;\; \mbox{if $x>0$},
\end{equation}
\begin{equation}
\delta \textbf{E} = (e C^2 x,0,0), \;\;\; \delta \textbf{B} = 0,
\end{equation}
where $C$ is a constant. In the following, in order to avoid
divergent values in the charge and energy calculation, we suppose
that $\delta \psi$ and $\delta E_x$  exist in large but finite
segment of $x$ axis: $-L/2 \leq x \leq L/2$. The constant $C$ is
related to the surface charge density by the following relation,
\begin{equation}
Q = \int^{L/2}_{-L/2} dx \,
    \delta {\bar{\psi}} \, \gamma^0 \,
    \delta \psi =
    \int^0_{-L/2} dx \,
    |\delta \psi_-|^2 +
    \int_0^{L/2} \! dx \, |\delta \psi_+|^2 =
    C^2 L.
\end{equation}
Considering the possibility of existence of stable topological
charged solution, we have to compare the density energy $\sigma_T$
of the solution having charge density $Q$ with the sum of the
energy of the kink (2.9) and the energy of charged non-topological
configuration (4.8)-(4.11),
\begin{equation}
\sigma_{NT} = \sigma_w + \sigma_{\delta \psi} + \sigma_{\delta E}.
\end{equation}
The energy density associated to $\delta \psi$ is
\begin{equation}
\sigma_{\delta \psi} =
                     m_0 \int^{L/2}_{-L/2}  dx \, |\delta \psi|^2 = m_0 Q,
\end{equation}
while the electric energy density is
\begin{equation}
\sigma_{\delta E} = \int^{L/2}_{-L/2}  dx \,
                    \frac{\delta E_x ^2}{2} \sim e^2 Q^2 L.
\end{equation}
There is stable topological charged solution if
\begin{equation}
\sigma_{T} < \sigma_{NT},
\end{equation}
that is when
\begin{equation}
Q \; \lesssim \; \frac{\omega - m_0}{e^2 (L - l)} \; \equiv \;
                 Q^{crit}.
\end{equation}
We observe that fermions from the broken phase (which have energy
$E>m_0$) can be captured by the wall giving rise to a charged
domain wall
\footnote{We are supposing that the wall is finite but the linear
dimension is much grater than the thickness of the wall. Moreover
the integration along the axis $x$ is performed over a distance
$L$ much greater than the linear dimension of the wall. So, we are
working in the hypothesis that $L>>l>>\Delta$.}.
Then when the charge density $Q$ due to the fermionic mode
solutions situated on the wall satisfy the inequality (4.17) we
have that the topological structure of a CDW does not decay in
domain wall and plane wave. In this way the existence of the CDW's
is connected with the critical charge density $Q^{crit}$. The
physical meaning of the critical charge density tell us that the
first configuration is energetically favorable if $Q<Q^{crit}$,
while if $Q>Q^{crit}$ we have a decomposition into a (kink)+(plane
waves).
%
%
%
%
%
%
\renewcommand{\thesection}{\normalsize{\arabic{section}.}}
\section{\normalsize{Conclusions and outlook}}
\renewcommand{\thesection}{\arabic{section}}
In this paper we have investigate the possible existence of
charged domain walls. An important difference from Q-balls is the
origin of the charge density $Q$. In our work the charge density
does not derive from a complex scalar field but it is due to the
presence of localized fermions  on the wall.
\\
We have analyzed the presence of an electric field (perpendicular
to the wall) and magnetic field (parallel to the wall). It will be
interesting to study as these fields might modify the interaction
of the charged wall with the surrounded plasma and how the Coulomb
barrier behaves in comparison with the absorption or repulsion
with the plasma.
\\
Moreover we have analyzed the stability of CDW's by confronting
two physical situations: kink with localized fermions on the wall
and (kink)+(fermion plane waves). The calculation of the energy
density, for both, allowed to find a condition for the existence
of CDW's with respect to the decomposition into a (kink)+(plane
waves). We have found there is a critical value for the charge
density, $Q^{crit}$: charged domain walls solutions exist when the
charge density $Q$ is smaller than $Q^{crit}$. In this case the
charged domain walls survive, otherwise the system divided into a
kink and plane waves, that is more favorable from an energetic
point of view.
\\
The fate of CDW's in the early Universe is determined by their
lifetime and if these defects actually exist in the Universe, they
may have interesting effects on the cosmology. An interesting
hypothesis consists in to consider such objects as they would have
survived until the present time and would contribute to the matter
density of the Universe as a form of charged dark matter. This
hypothesis has just been considered as regards the Q-balls, that
are the ground state configurations for fixed charge $Q$ in
theories with interacting scalar fields that carry some global
U(1) charge \cite{{KUSENKO1},{KUSENKO2}}. It also will be
interesting to understand the effect of CDW's on the structure
formation in the early Universe.
\\
However it must be pointed out that the existence of domain wall
is still questionable. Indeed, in general the gravitational
effects of just one such wall stretched cross the Universe would
introduce a large anisotropy into the relic blackbody radiation.
So that CDW's could have survived until today in the form of
bubbles of radius R. In this case, it turns out that the bubbles
get stabilized by the Coulomb repulsion. Indeed, the authors of
Ref.[3] showed that these defects occur whenever the total charge
is greater that a certain minimum value. In this case we are left
with a gas of charged domain walls which, indeed, could be
cosmological important.
\\
Another important comment regards the connection between the
primordial magnetic field and the magnetic field on the plane of a
CDW. We can estimate its strength supposing that
$B \sim B^{(max)}_z = eQ/2 \sim eQ^{crit}$.
Taking, for example, $\omega = 2m_0$ and $L=10 l$ in Eq.~(4.17) we
have
$Q^{crit} \sim m_0 /l$,
and for the magnetic field
\begin{equation}
B \sim \frac{g_Y \eta}{e \, l}.
\end{equation}
Taking $g_Y/e \sim 1$ and $l = 10 \Delta$ we obtain, at the
electroweak phase transition ($v \sim 10^2 \, \mbox{GeV}$), $B
\sim 10^{22}\, \mbox{Gauss}$. This value of the magnetic field is
in accordance with the estimate obtained at the electroweak phase
transition for primordial magnetic field \cite{DOLGOV}.
\\
Another consideration regards the possibility that CDW's can decay
after the electroweak phase transition. They could protect baryons
from the erasure of baryon number due to the sphaleron
transitions, therefore they can also create the barionic asymmetry
in the early Universe \cite{DONALD}. In any case, it is beyond the
aim of present paper to examine these problems, which will be
object of an upcoming work.
%
%
%
%
%
%
\\
\\
\vfill
\newpage
\renewcommand{\thesection}{\normalsize{\arabic{section}.}}

\vfill
\newpage
\end{document}